%% file: main.tex
\documentclass{article}
\usepackage[utf8]{inputenc}
\usepackage{graphicx}
\usepackage[portrait, margin=0.8in]{geometry}
\usepackage{hyperref}
\usepackage{authblk}
\usepackage{todonotes}
\usepackage{enumitem}
\usepackage{siunitx}
\usepackage{amsmath,amsthm,amssymb}
\usepackage{braket}
\usepackage[sorting = none, backend = bibtex, style=numeric-comp]{biblatex}

\title{An Efficient All-to-All GCD Algorithm for Low Entropy RSA Key Factorization}

\author[1]{Elijah Pelofske\thanks{Email: elijah.pelofske@protonmail.com}}
\affil[1]{Independent}

\date{\vspace{-6ex}}

\begin{document}
\maketitle

\begin{abstract}
\input{abstract}
\end{abstract}

\input{text}

\end{document}

%% file: abstract.tex
RSA is an incredibly successful and useful asymmetric encryption algorithm. One of the types of implementation flaws in RSA is low entropy of the key generation, specifically the prime number creation stage. This can occur due to flawed usage of random prime number generator libraries, or on computers where there is a lack of a source of external entropy. These implementation flaws result in some RSA keys sharing prime factors, which means that the full factorization of the public modulus can be recovered incredibly efficiently by performing a computation GCD between the two public key moduli that share the prime factor. However, since one does not know which of the composite moduli share a prime factor a-priori, to determine if any such shared prime factors exist, an all-to-all GCD attack (also known as a batch GCD attack, or a bulk GCD attack) can be performed on the available public keys so as to recover any shared prime factors. 
This study describes a novel all-to-all batch GCD algorithm, which will be referred to as the \emph{binary tree batch GCD} algorithm, that is more efficient than the current best batch GCD algorithm (the remainder tree batch GCD algorithm). A comparison against the best existing batch GCD method (which is a product tree followed by a remainder tree computation) is given using a dataset of random RSA moduli that are constructed such that some of the moduli share prime factors. This proposed binary tree batch GCD algorithm has better runtime than the existing remainder tree batch GCD algorithm, although asymptotically it has nearly identical scaling and its complexity is dependent on how many shared prime factors exist in the set of RSA keys. In practice, the implementation of the proposed binary tree batch GCD algorithm has a roughly $6x$ speedup compared to the standard remainder tree batch GCD approach. 

%% file: text.tex
\section{Introduction}
\label{section:introduction}

RSA \cite{10.1145/359340.359342}, named by the acronym formed from the first letter of the last names of the inventors, is an asymmetric cryptosystem that enables secure and efficient information exchange. The security of RSA comes from the hardness of integer factorization. Briefly, the RSA system is defined as follows:

\begin{enumerate}[noitemsep]
    \item Choose two prime numbers $p$ and $q$ at random. Large prime numbers can be reliably chosen using very fast probabilistic primality test algorithms. 
    \item Multiply them together to form the \emph{public key modulus} $N = p \cdot q$
    \item Compute the Euler totient function $\phi(N) = (p-1)\cdot(q-1)$. The Euler totient function denotes how many integers less than $N$ (and greater than $0$) are also coprime to $N$ (meaning that no factors are shared with with $N$). 
    \item Choose an integer $e$ such $e$ is coprime to $\phi(N)$. A typically choice is a small prime number, such as 65537. 
    \item Compute $d$ such that $e \cdot d \equiv 1 \mod \phi(N)$
    \item The pair of positive integers $(e, N)$ is the public key. $n$ is referred to as the \emph{modulus}
    \item The pair of positive integers $(d, n)$ is the private key. 
\end{enumerate}

A message is encrypted by $\text{ciphertext} = \text{message}^e \mod N$, and then can be decrypted by $\text{message} = \text{ciphertext}^d \mod N$

In order for an attacker to obtain the private key, they must know what $\phi(N)$ is, and generally that requires knowing what $p$ and $q$ are. RSA as an encryption scheme has been incredibly successful for a long period of time. However, there are a number of very specific parameter choices that can cause a particular instance of an RSA cryptosystem to be vulnerable to attack; these include certain choices of the public exponent, having a private key that is too small \cite{boneh1999cryptanalysis}, or even $p$ and $q$ being too close together \cite{cryptoeprint:2009/318, cryptoeprint:2023/026}. The most basic attack is to simply attempt to compute the prime factorization of $N$ \cite{cryptoeprint:2020/697, cavallar1999factorization, cowie1996world} - and to combat this, $p$ and $q$ are typically chosen to be very large (on the order of $1024$ bitlength), which makes factorization of $N$ intractable for the state of the art factorization algorithms and computing hardware.

The focus of this study however is a different attack that is remarkably simple and that is typically known as a \emph{batch GCD attack}, which was first described and utilized (independently and concurrently) in refs. \cite{180213, Lenstra2012RonWW}. Since then, batch GCD attacks have been used in a number of studies in an effort to evaluate the security of RSA keys on the internet, specifically for the incredibly simple flaw of repeated usage of prime factors \cite{10.1145/2987443.2987486, 7284337, janovsky2020biased, barbulescu2016rsa, cryptoeprint:2013/599, defcon_29_batch_gcd, github_defcon29_batch_gcd, defcon_26_batch_gcd, raddumfactoring, kumar2017parallelized, 10.1145/3627106.3627120}, including as part of security evaluations of commercial products \cite{proton_batch_GCD, research_blog_factoring_RSA_moduli}. The overall state of batch GCD analysis on public RSA keys, from these previous studies, is that it remains a persistent issue - although the percentage of RSA moduli that have this flaw are generally quite low. The vulnerability that the batch GCD attack can exploit is that the Greatest Common Divisor (GCD) algorithm, also known as Euclid's algorithm, allows the largest shared factor between two integers to be very efficiently computed -- and therefore if any prime factor is shared between two RSA public moduli, a GCD operation between those two moduli would quickly give one of the prime factors, thus breaking the security of that implementation of RSA\footnote{Note that the Euler totient function computation in RSA can easily allow for more prime factors to be used than $2$. A batch GCD attack can still be applied to RSA keys that have more than one prime factor -- there is just the possibility that only a portion of the prime factors could be recovered. }. The reason that the attack is referred to as a \emph{batch} GCD attack is because one must compute the greatest common divisor between all pairs of integers (in this case, specifically RSA moduli) in order to comprehensively find any shared prime factors (in particular since it is not known a-priori which RSA moduli share prime factors, if any)\footnote{We will interchangeably use the terms \emph{bulk GCD}, \emph{batch GCD}, and \emph{all-to-all GCD}}. In the brute-force approach of a batch GCD attack, a separate GCD computation must be performed for each pair of RSA moduli, which corresponds to a fully connected graph of GCD comparisons, meaning in total $\frac{M \cdot (M-1)}{2}$ GCD computations must be performed where $M$ is the total number of RSA keys being analyzed. This brute force approach, while efficient because the individual GCD computations are efficient, in practice requires an immense amount of computation time if $M$ is sufficiently large - for example at the scale of RSA keys in use in information technology systems today. The fundamental flaw that a batch GCD attack exploits is an underlying implementation flaw where identical prime numbers are inadvertently being generated independently on (likely on entirely different computers), perhaps due to a lack of sources of entropy for seeding a pseudo-random-number generator on the hardware at the time the keys are generated\footnote{There is an ironic case of low-entropy RSA keys with shared prime factors that are not factorable by a GCD attack; this is the case where two or more RSA moduli share exactly the same prime factors due to an implementation flaw or low entropy (in other words, they are exactly equal to each other). Since the greatest common divisor of this number is simply the modulus, the task of breaking such RSA keys can not be efficiently addressed using GCD. The only way for batch GCD to find one of the prime factors is if that prime factor exists elsewhere in a different RSA modulus that is being analyzed in the same batch GCD run. }. Ref. \cite{180213} investigates various causes of low entropy and implementation flaws causing the same prime numbers to be produced on certain computers.

Although RSA is the focus of the batch GCD attacks, due to the inherent necessity of generating unique prime numbers, there exist other cryptosystems that require the creation and usage of large prime numbers as part of the algorithm. Namely, ElGamal (uses 1 prime), Diffie-Hellman (uses 1 prime number), and DSA (which uses 2 prime numbers). Public keys and public configurations of these cryptosystems can then also be analyzed in a batch GCD computation, or likely even easier by enumerating over a set of RSA moduli to find if any of the primes are evenly divisible into a modulus, to potentially find any shared prime factors due to implementation flaws or limited entropy sources.

The proposed batch GCD algorithm will be compared against the current most efficient batch GCD algorithm (that was used in ref. \cite{180213} for example), and that is described in ref. \cite{bernstein2004find}. We will refer to this algorithm as the \texttt{Remainder Tree Batch GCD} algorithm. The overall structure of the remainder tree GCD algorithm is to first form a product tree of the input RSA moduli, and then to construct a remainder tree, from which GCD operations at the leaf level of the tree produce shared factors found in the set of RSA keys. 

For consistency throughout the text and plots, we will refer to the novel batch GCD algorithm that this study proposes as the \texttt{Binary Tree Batch GCD} algorithm, named for the central use of performing GCD operations on the binary product tree when it is being constructed so as to obtain the combined shared factors throughout the entirety of the set of RSA moduli.

\subsection{Terminology and Variable Definitions}
\label{section:introduction_variable_defs}

\begin{itemize}[noitemsep]
    \item $N$ denotes the set of integers on which we are performing GCD operations. Therefore, in particular these are the list of public key moduli that are composite (semiprime) numbers. Therefore, indexes of $N$ denote elements of that array; such as $N_1$. Note that in Section~\ref{section:introduction} $N$ is used to denote a single public key modulus; this is to make the conventions of the RSA algorithm consistent, but for the remainder of the paper $N$ will denote the set of public key moduli. 
    \item $M$ denotes the size of $N$. For example, $M=1000$ means that there are $1000$ integers that we wish to perform all-to-all GCD computations on. 
    \item $B$ denotes the integer that is the product of all of the non trivial common divisors found during the all-to-all GCD computation. 
    \item $n$ denotes the bit size of an integer. In the case of RSA public moduli, this will generally be some large power of $2$, such as $1024, 2048, 4096, 8192$. This size is important for determining the overall runtime of the algorithm. For simplicity of the analysis, and consistency with the existing literature, we will assume that $n$ is the same for the all of the integers. 
    \item $C$ denotes the total number of shared factors that exist within all of the $N$ integers. 
\end{itemize}

\subsection{Time Complexity Assumptions}
\label{section:introduction_time_complexity_assumptions}
For the asymptotic computational complexity of all to all GCD algorithms, there are several fundamental computations that we need to define the complexity of before building up to the complexity of the larger algorithms. 

The first is multiplication. We will assume that FFT product tree multiplication \cite{bernstein2008fast} is being used, since this works quite well for taking the product of a large quantity of large integers. Note that there has been an incredibly fast integer multiplication algorithm discovered \cite{harvey2021integer} that has $\mathcal{O}(n \log n)$, but here we assume that the multiplication complexity of two $n$ bit integers (using FFT multiplication) is $\mathcal{O}(n \log n \log \log n)$ \cite{bernstein2008fast}. We will assume that all of the integers are $n$-bit integers. For a product tree, there are at most $1 + \log_2 M$ levels of the binary tree. Therefore, the complexity of the product tree computation is $\mathcal{O}( nM \log M \log nM \log \log nM )$

We will assume that the binary GCD$(a, b)$ operation has asymptotic computational complexity of $\mathcal{O}(\log(a \cdot b)^2)$ for arbitrarily large integers (which applies when considering RSA public key moduli), where $a \cdot b$ denotes the total bits that are input to the GCD computation \cite{Sunar2005, bach1996algorithmic}. Note however that like integer multiplication, there are new improved implementations of GCD \cite{cryptoeprint:2019/266} that could have even better scaling. 

These complexity assumptions are used to compare the binary tree batch GCD algorithm against the existing product tree and remainder tree batch GCD algorithm, and because these time complexity assumptions are likely reasonable for current standard library implementations of these algorithms. However, as already mentioned, there are algorithmic improvements both to integer multiplication and GCD that would allow for even better asymptotic computational complexity scaling.

\begin{figure*}[tbh!]
    \centering
    \includegraphics[width=0.99\textwidth]{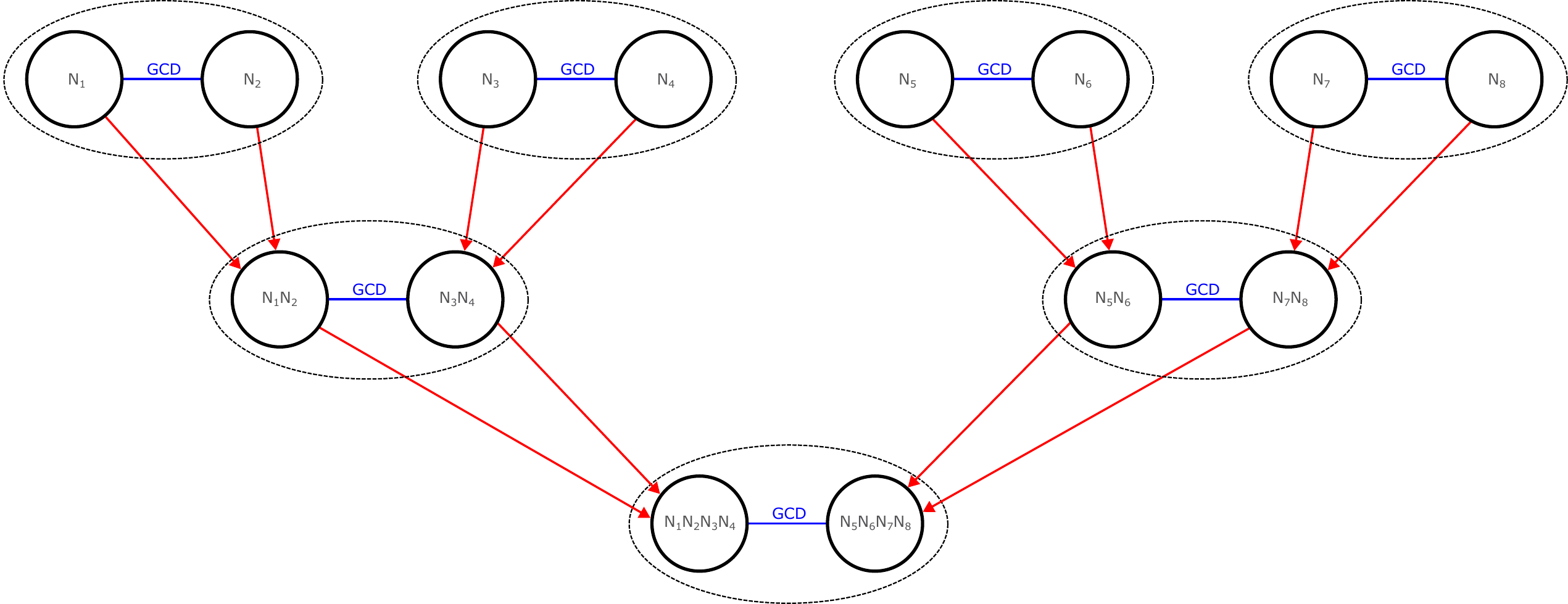}
    \caption{Diagram of the all-to-all GCD algorithm for $8$ integers labeled $N_1, \ldots N_8$. Blue lines denote the GCD operation, red arrows denote the integer multiplication operation, black circles denote integers, and the thin black ovals denote the logical pairs of integers that have a GCD operation performed on them and are then multiplied together. Because this example uses exactly $8$ integers, this is a complete binary tree. The algorithm starts at the leaf nodes, which are pairs of the input integers. A single GCD operation occurs between each pair, and then each pair is multiplied together (denoted by the two red arrows). This operation is performed recursively as the tree is built until the final step where the GCD between the product of one half of the integers and the other half of the integers is performed. Note that at that final step, there is no need to compute the product of those two large integers. In total, the number of GCD operations is exactly equal to the number of nodes in the binary tree (which in this diagram are represented as meta-groupings of nodes shown by the ovals), which is equal to the number of nodes in the binary tree minus $1$. The key characteristic of this recursive GCD approach is that combined all of the individual GCD computations have covered the entirety of the $\frac{M \cdot (M-1)}{2}$ edges in the complete all-to-all graph formed by the clique of the set of $N$ integers (where each edge on this conceptual clique is a GCD operation). This can be seen at the root of this binary tree, where that single GCD operation has covered $16$ individual GCD($N_i, N_j$) operations -- the output is of course not the distinct common divisors, but is instead the \emph{product} of all of the common divisors between those two large integers. At each GCD operation in the binary tree, the result we get is either $1$ or a non-trivial greatest common divisor. If the GCD is a non trivial common divisor, then we record it - and specifically multiply all of the found non trivial common divisors together into a single integer denoted as $B$. $B$ is the thus the product of all of the shared factors within the set of input integers $N$.  }
    \label{fig:GCD_binary_tree}
\end{figure*}

\section{Binary Tree Batch GCD Algorithm}
\label{section:algorithm_description}

The Binary Tree Batch GCD algorithm is defined by the following three steps:

\begin{enumerate}[noitemsep]
    \item Construct a product tree out of the integers in $N$ where at each node of the tree we take the GCD between the two integers. 
    \item Take all of the GCD outputs (which is equal to $n-1$ integers) and take the product of those integers (also using a product tree, for efficiency especially if the number of non-trivial divisors is large). Call this integer $B$. If $B$ is equal to $1$, then there are no shared prime factors in $N$. 
    \item Take the GCD between each integer in $N$ and $B$. Any non $1$ outputs from this GCD enumeration are shared factors. 
\end{enumerate}

The central idea of this algorithm is based on the very nice properties of the GCD algorithm. Namely that if we want to compute the GCD between all pairs of a given set of input integers, we can extract the product of half of those GCD computations by splitting the integer set in half and performing GCD on the product of both halves. This does not cover the entirety of the complete graph, but it does cover a very large number of them. We can however use a binary tree to recursively find the GCD between all pairs of integers; this is shown in Figure~\ref{fig:GCD_binary_tree}. Recursively performing this GCD operation while also building a product tree allows us to find the product of all common divisors in the all-to-all GCD computation. We call this integer $B$ (note that if $B=1$, there are no shared factors in $N$). This process can be formulated as the GCD operations at the nodes of this binary product tree covering all edges in an undirected clique graph of size $M$; thus performing an all-to-all GCD computation. Next, in order to extract the actual factors, we need to enumerate through every integer in $N$ and perform a GCD operation between $B$ and each integer in $N$. This will recover all shared factors in $N$. The exact number of GCD operations required to do this is exactly $M + M-1$. 

There is a potential complication that may arise however. If one or more of the integers in $N$ contains two (or more) factors that are also factors in other integers in $N$, then while we are computing the GCD between $B$ and each element of $N$ at some point we will get a GCD output that is equal to the integer $N_i$ that we just fed as input the GCD algorithm along with $B$. This case means that the common divisors of $N_i$ will be found elsewhere in the GCD list iteration, but it will take some more operations to find what those factors are (the simplest would be to take the modulo of $N_i$ by the other found factors). In other words, in order to uniquely identify what RSA key contained multiple shared factors, slightly more computation time is required.

\begin{itemize}[noitemsep]
    \item The time complexity of the product tree construction of step 1 is $\mathcal{O}( nM \log M \log nM \log \log nM )$. Note that this is exactly the same as the product tree construction used in \emph{remainder tree GCD} \cite{bernstein2008fast}. The one slight difference is that the final large product is not formed (as illustrated by Figure~\ref{fig:GCD_binary_tree}), but this does not change the overall asymptotic computational complexity scaling. 
    \item The time complexity of the GCD operations in step 1 is lower bounded by  $\Omega( \log ( (n \cdot \frac{M}{2})^2 )^2 )$, and is upper bounded by $\mathcal{O}( \log M \log ( (n \cdot \frac{M}{2})^2 )^2 )$. The lower bound comes from the fact that at least that number of bit operations needs to be performed at the root of the binary tree (shown in Figure~\ref{fig:GCD_binary_tree}). This is because at that root node, we perform one GCD operation that is between the product of the two halves of the RSA moduli set; each of those two halves is size $n \cdot \frac{M}{2}$. But, for the other levels of the binary tree smaller GCD operations also occur (at each of the$\log M$ levels of the binary tree) - and therefore the upper bound is found by assuming a full $\log ( (n \cdot \frac{M}{2})^2 )^2$ bit operations will happen at every level of the binary tree for the GCD comparisons. In the algorithm, fewer bit operations are required closer to the leaf nodes, and so this bound is not precise but is instead an upper bound. 
    \item The time complexity of step 2 of the algorithm is $\mathcal{O}( nC \log C \log nC \log \log nC )$. Note that $n \cdot C$ denotes the total bits being used in the computation. 
    \item The time complexity of step 3 of the algorithm is $\mathcal{O}(M \cdot \log(n \cdot C \cdot n)^2)$. Note that the complexity is dependent on how many shared factors there are; the larger $C$ is, the more compute time this will require. 
\end{itemize}

This assumes that there are no cases as described above where there exist repeated common factors (for example where an RSA modulus contains two distinct primes, both of which are shared elsewhere in the overall set of RSA keys). Such cases do add real computation time in the form of modulo operations using very large numbers (or, in whatever other way repeated shared prime factors are handled by the implementation).

Compared to the remainder tree algorithm of ref. \cite{bernstein2004find}, the binary tree batch GCD binary has (effectively) the same asymptotic computational complexity scaling -- which in particular is defined by the computation of the product tree. In other words, assuming both algorithms use the same product tree implementation, the asymptotic computational complexity scaling is essentially the same. However, the binary tree batch GCD algorithm does not have the cost of computing the remainder tree (which has the same asymptotic scaling as the product tree). The remainder tree batch GCD algorithm \cite{bernstein2004find} has $4$ distinct steps to it, each of which have a worst-case scaling of $\mathcal{O}(b \log(b)^2 \log \log b)$ where $b$ is the total number of bits being used. This new batch GCD algorithm has the same worst case scaling (caused by the product tree computation), but saves in computational complexity when the total number of shared factors is small. 

\begin{figure*}[th!]
    \centering
    \includegraphics[width=0.32\textwidth]{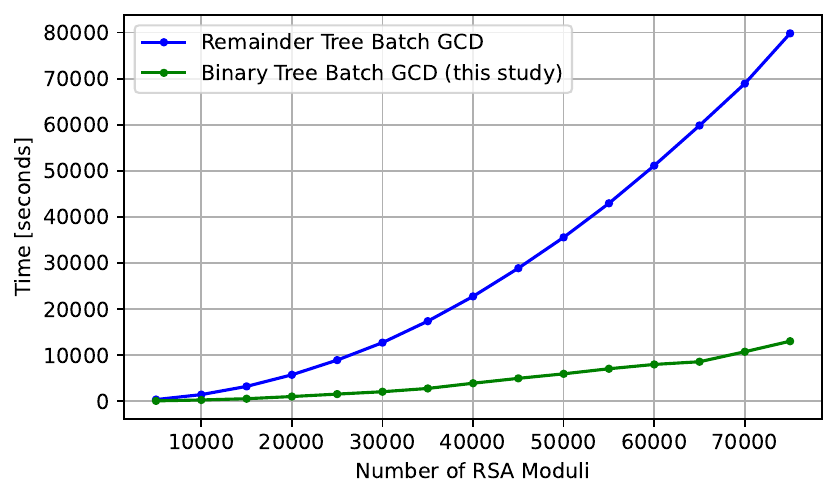}
    \includegraphics[width=0.32\textwidth]{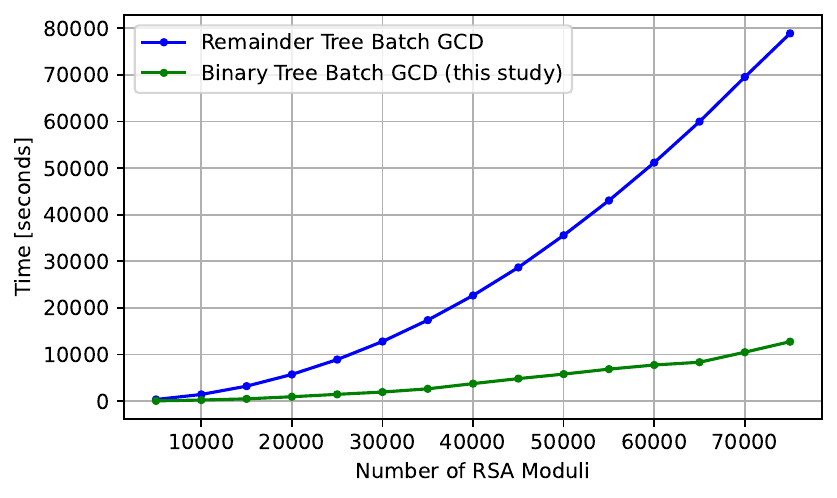}
    \includegraphics[width=0.32\textwidth]{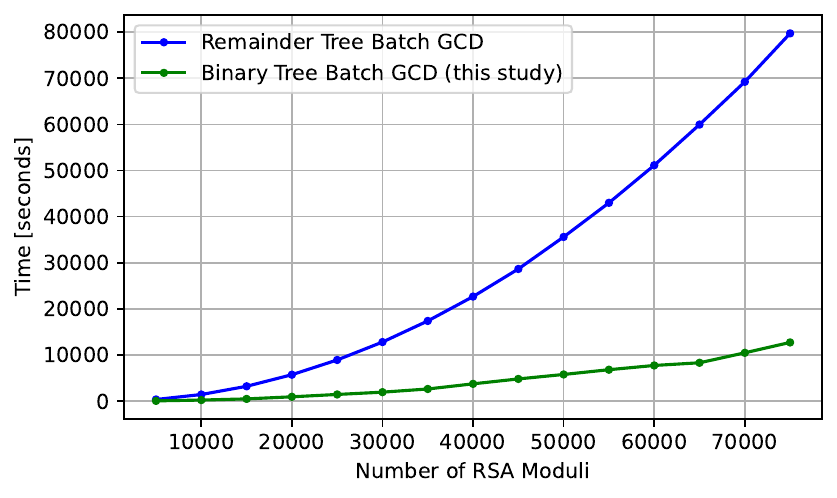}
    \includegraphics[width=0.32\textwidth]{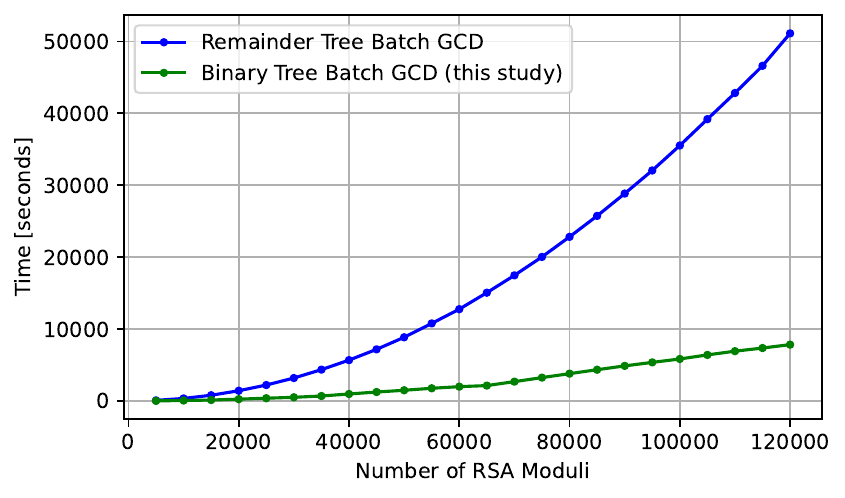}
    \includegraphics[width=0.32\textwidth]{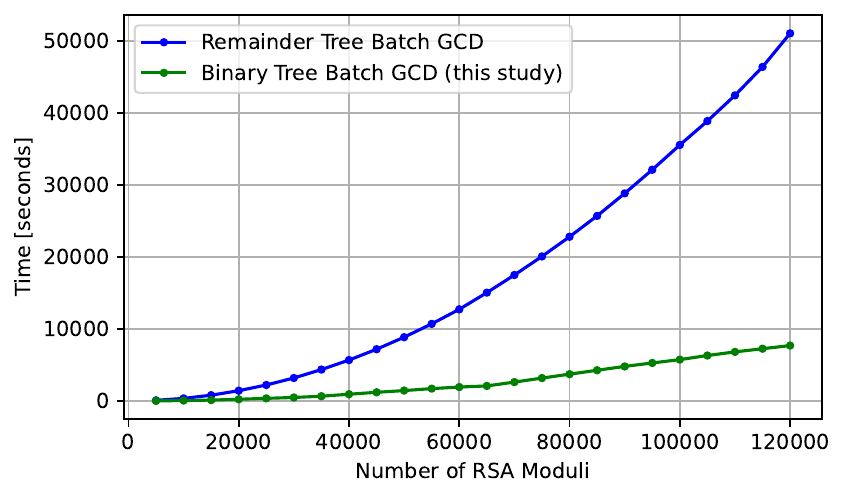}
    \includegraphics[width=0.32\textwidth]{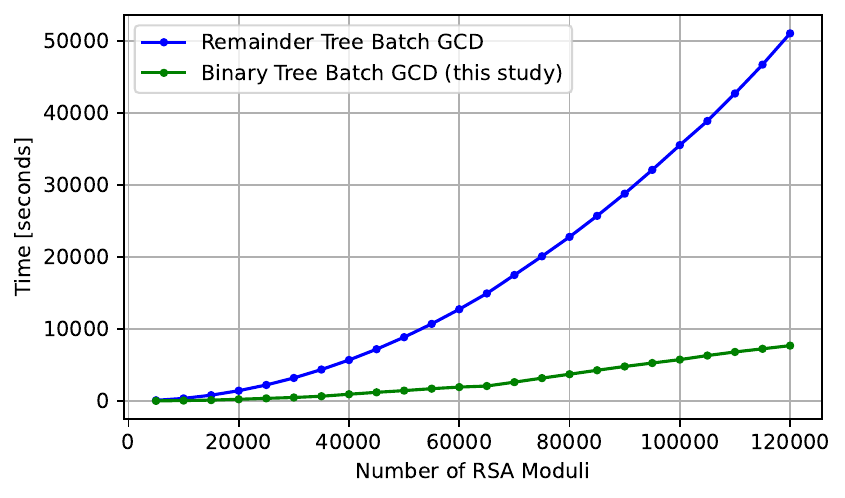}
    \caption{Compute time scaling for both the original remainder tree batch GCD algorithm (blue), and the binary tree batch GCD approach proposed in this study (green). The top row used $2048$ bit RSA moduli, and the bottom row used $1024$ bit RSA moduli. The data shown in the left column had exactly $2$ weak RSA moduli ($2$ RSA keys shared a prime factor) in the pool of RSA keys, the plots in the middle column show the results for exactly $100$ weak RSA moduli, and the right column is for exactly $1000$ weak RSA moduli. The binary tree batch GCD scaling is consistently better than the remainder tree batch GCD algorithm. }
    \label{fig:timing_scaling}
\end{figure*}

\section{Computational Complexity Timing Results}
\label{section:timing_results}
In order to measure the compute time scaling of both the remainder tree batch GCD algorithm and the batch GCD algorithm presented in this study, we measure the CPU time (specifically the \emph{CPU process time}) required to correctly identify all of the shared prime factors in a set of RSA public moduli. So as to make the configuration of tested RSA keys reasonably simple, and also small enough to be evaluated without HPC resources, we define a set of RSA moduli some of which are intentionally constructed to have shared prime factors with other RSA public moduli. There are two RSA key configurations that will be created and subsequently used in the performance simulation analysis for the two batch GCD algorithms; RSA moduli key size of $1024$ and $2048$ bits, and the number of weak RSA keys, meaning how many of the RSA keys share a single unique prime factor. This set of RSA keys is created such that the weak RSA keys are comprised of pairs of keys that share exactly one prime factor. This means that there are no repeated shared prime factors. All RSA keys used in these simulations have exactly two prime factors. The scaling measurements that are performed are done in terms of the complexity scaling as a function of how many RSA moduli are in the set $N$ that are being analyzed for potentially sharing primes with each other; specifically, this means that the scaling analysis is assuming a fixed number of those RSA moduli have shared prime factors. 

The implementations are written in Python 3 code, and the full code and dataset of random RSA moduli is publicly available on Github\footnote{\url{https://github.com/epelofske65537/binary_tree_Batch_GCD}}. All simulations are performed on a \texttt{Apple M1 Max} device. The implementation for the algorithm of ref. \cite{bernstein2004find}, which is comprised of a product tree followed by a remainder tree, is Python code given in ref. \cite{batchgcd_remainder_tree_python}. This implementation timing analysis is intended to directly compare the existing remainder tree approach, and the binary tree batch GCD approach proposed in this study -- in particular by using the same libraries and basic math functions. Therefore, this timing analysis does not necessarily use implementations that adhere to the time complexity assumptions (in particular for integer multiplication) outlined in Section~\ref{section:introduction_time_complexity_assumptions}. The exact RSA moduli used in the simulations is also fixed between the two algorithms by using the same random seed. 

Simulations are performed starting at $5000$ RSA moduli, up to $75,000$ RSA moduli for $2048$ bit keys, and up to $120,000$ RSA moduli for $1024$ bit keys.

\begin{figure*}[th!]
    \centering
    \includegraphics[width=0.49\textwidth]{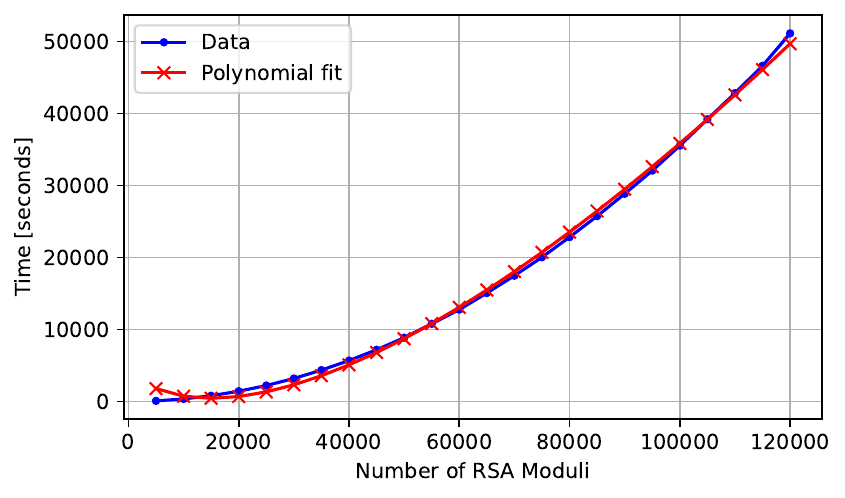}
    \includegraphics[width=0.49\textwidth]{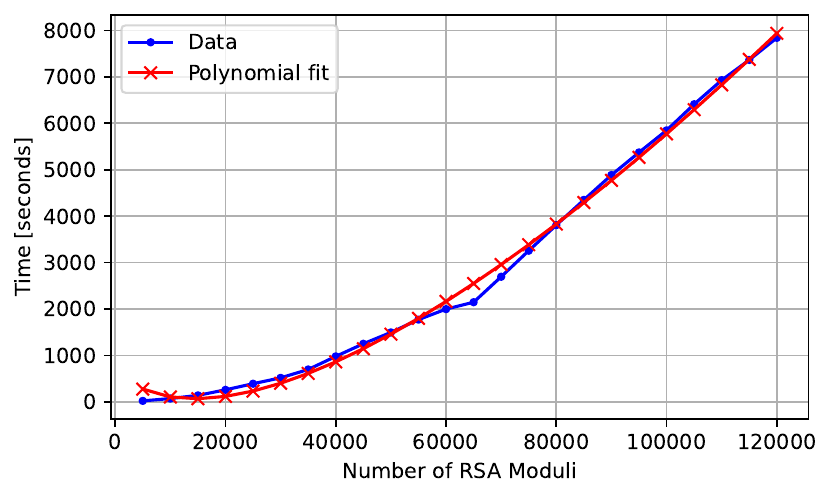}
    \caption{Polynomial curve fit scaling, in terms of CPU compute time, for $1024$ bit RSA keys with exactly $1000$ weak keys. Left plot shows the scaling for the remainder tree batch GCD algorithm, and the right plot shows the scaling for the binary tree batch GCD algorithm. }
    \label{fig:curve_fit_plots}
\end{figure*}

\begin{table*}[ht!]
\centering
\begin{tabular}{ |p{10cm}|c||c|c| }
 \hline
 Algorithm and Settings  & $a$ & $b$ & $c$  \\ 
 \hline
 \hline
 \hline
 Remainder Tree Batch GCD - $1024$ bit keys - $2$ weak keys & $1.10404$ & $-3.00178$ & $4686$  \\ 
 \hline
 Remainder Tree Batch GCD - $1024$ bit keys - $100$ weak keys & $1.10364$ & $-2.9866$ & $4609.3$  \\ 
 \hline
 Remainder Tree Batch GCD - $1024$ bit keys - $1000$ weak keys & $1.10406$ & $-3.00211$ & $4680.2$  \\ 
 \hline
 Remainder Tree Batch GCD - $2048$ bit keys - $2$ weak keys & $1.15741$ & $-4.91505$ & $7692.1$  \\ 
 \hline
 Remainder Tree Batch GCD - $2048$ bit keys - $100$ weak keys & $1.15673$ & $-4.87067$ & $7490.6$  \\ 
 \hline
 Remainder Tree Batch GCD - $2048$ bit keys - $1000$ weak keys & $1.15722$ & $-4.90262$ & $7639.6$  \\ 
 \hline
 \hline
 Binary Tree Batch GCD - $1024$ bit keys - $2$ weak keys & $1.03531$ & $-1.45274$ & $777.6$  \\ 
 \hline
 Binary Tree Batch GCD - $1024$ bit keys - $100$ weak keys & $1.03528$ & $-1.45218$ & $776.2$  \\ 
 \hline
 Binary Tree Batch GCD - $1024$ bit keys - $1000$ weak keys & $1.03536$ & $-1.45255$ & $780.9$  \\ 
 \hline
 Binary Tree Batch GCD - $2048$ bit keys - $2$ weak keys & $1.06432$ & $-1.91702$ & $1270.1$  \\ 
 \hline
 Binary Tree Batch GCD - $2048$ bit keys - $100$ weak keys & $1.06431$ & $-1.91628$ & $1267.5$  \\ 
 \hline
 Binary Tree Batch GCD - $2048$ bit keys - $1000$ weak keys & $1.06443$ & $-1.91562$ & $1278.6$  \\ 
 \hline
\end{tabular}
\caption{Least-squares curve fitting results for the polynomial $x^a + bx + c$, where $a, b, c$ are the constants to be fitted. The data the polynomial is fitted to is the measured CPU time for both algorithms; this is then repeated for both $1024$ and $2048$ bit RSA moduli, with $2$, $100$, and $1000$ weak moduli (due to shared prime factors). Polynomial coefficients are rounded to 5 decimal places.  }
\label{table:scaling_fits}
\end{table*}

Figure~\ref{fig:timing_scaling} plots the exact CPU time usage scaling for both batch GCD algorithms, under two different varied settings; number of weak RSA keys, and bitlength of the RSA keys. For all simulations the compute time of the binary tree batch GCD approach is better. So as to determine what the computational complexity scaling is of these timing results, a least-squares curve fit of the data to the polynomial $x^a + bx + c$ is performed using the Python 3 library scipy \cite{2020SciPy-NMeth}. Here, $a, b, c$ are the real coefficients that are fitted. This same curve fitting is applied to both algorithms, and for all settings\footnote{Note that because of the nature of the curve fit procedure, it is intended to be a good fit for computing the scaling for large inputs to the algorithms, but are not good fits for small inputs (specifically less than about $5,000$ RSA moduli). }. Table~\ref{table:scaling_fits} gives the exact fitted polynomial coefficients, and Figure~\ref{fig:curve_fit_plots} gives a polynomial curve fit examples for one of the RSA key settings settings for both algorithms. The fitted exponent $a$ is smaller for all of the binary tree batch GCD data compared to the remainder tree batch GCD data.

\begin{figure*}[th!]
    \centering
    \includegraphics[width=0.49\textwidth]{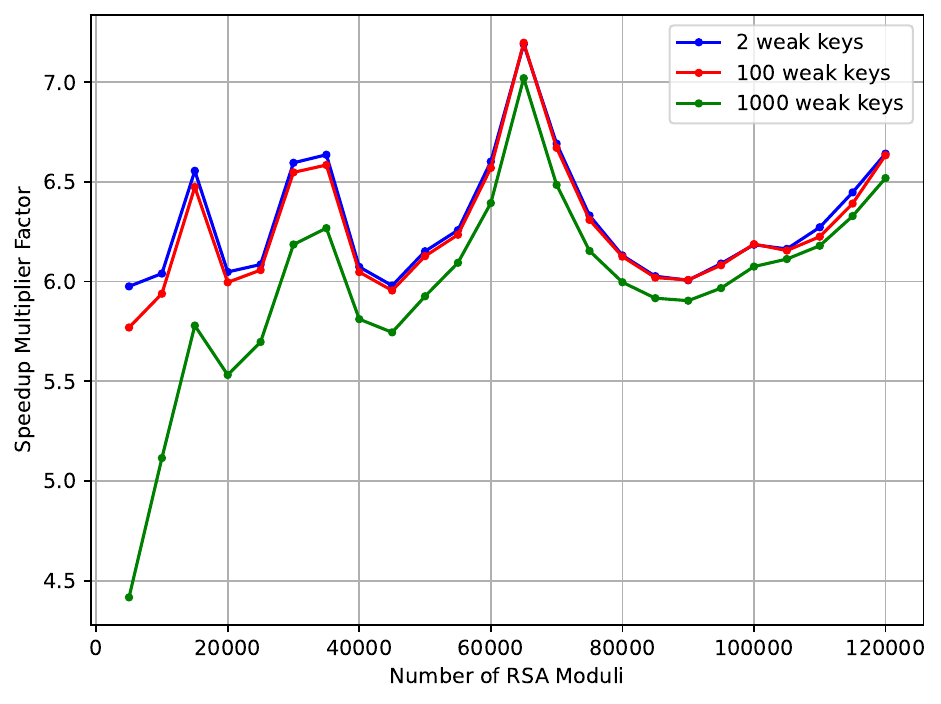}
    \includegraphics[width=0.49\textwidth]{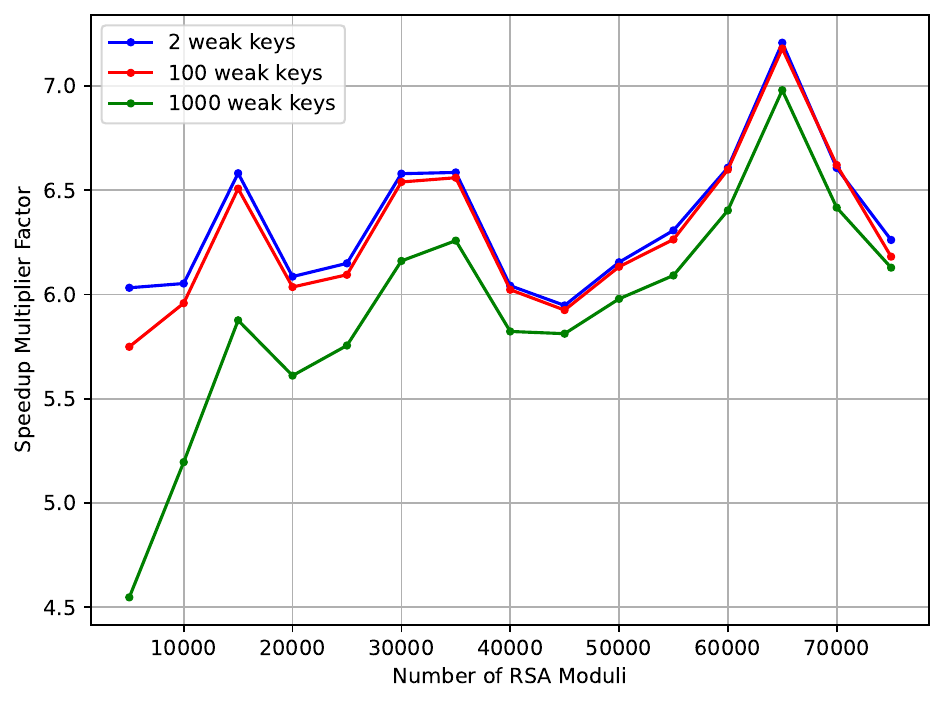}
    \caption{Scaling of the multiplier of how much faster the binary tree batch GCD run was compared to the remainder tree batch GCD run, as a function of the number of RSA moduli being analyzed by the batch GCD algorithm. }
    \label{fig:multiplier_scaling}
\end{figure*}

Another important scaling analysis is the ratio of the compute time used between these two algorithms, and how that time scales as the number of RSA moduli being analyzed increases. This analysis is shown in Figure~\ref{fig:multiplier_scaling}, where the multiplicative factor for how much faster the binary tree batch GCD compute was compared to the remainder tree batch GCD compute. The multiplicative factor is not very consistent as a function of the number of RSA moduli, although it has a weakly increasing trend suggesting that the binary tree batch GCD could have a non-constant scaling advantage over the remainder tree batch GCD algorithm that is dependent on system size (where system size is how many RSA moduli are being analyzed for potential shared prime factors). However, for the scale at which these computations were performed, Figure~\ref{fig:multiplier_scaling} shows that on average the binary tree batch GCD algorithm is approximately $6x$ faster than the remainder tree batch GCD algorithm. The high variability in the ratio between compute times is likely due to various tradeoffs that occur in the Python 3 library implementations being used as the integers progress to larger and larger sizes.

\section{Discussion and Conclusion}
\label{section:conclusion}

This study proposed an algorithmic improvement over the existing batch GCD algorithms. Based on the scaling analysis performed, this proposed binary tree batch GCD algorithm will allow for larger scale batch GCD analysis of RSA, and other public cryptosystem, keys - allowing weak low entropy keys with shared prime factors to be factored more efficiently. 

There are several further optimizations that could be made to this binary tree batch GCD algorithm, which were not explored in the current study. Generally, these optimizations would likely give marginal improvement of the algorithm. 

\begin{enumerate}[noitemsep]
    \item Reduce the size of $B$ during the GCD enumeration in Step 3 of the binary tree batch algorithm whenever a non trivial divisor is found by dividing out the found factor. This would decrease the computational cost of subsequent GCD operations, but the computational cost would be dividing out the found divisor from $B$. 
    \item The product tree that forms the integer $B$ could likely be optimized so that the multiplications that take place are more balanced compared to naively applying the product tree to all non-trivial divisors. The reasoning for this is that integer multiplication is less efficient when the two integers are dramatically different sizes, which is why the product tree technique works so well. The computed divisors from the GCD operations in the binary tree could be of dramatically varying size, and therefore one could optimize the product tree multiplication to make each product more balanced, rather than less balanced. 
    \item If any of the found divisors during the binary tree stage are prime numbers (or, just one of the RSA moduli in the edge case where all prime factors of that modulus occur more than once in the set of RSA keys), then they would not need to be multiplied into $B$, which saves on the computation to construct $B$, but also saves on the subsequent GCD enumeration in Step 3 since $B$ will be smaller. The cost of checking this would be running a primality test for all non trivial divisor outputs that appear to be the correct bitlength, which may be too high of a cost to benefit the saved computation time. 
    \item Similar to the previous item, if any of any non $1$ divisors are found at the leaf level of the binary tree stage, they are necessarily either a prime factor or one of the RSA moduli (if all of the prime factors of that modulus occur more than once in the set of RSA keys). In either case, they do need to be multiplied into $B$. This does not require primality test checking, so this case is an essentially free optimization. However, such cases will not occur very frequently since the number of \emph{GCD edges} covered by the leaf nodes is small compared to the bulk of the GCD operations. 
\end{enumerate}

The timing results are reported from using a Python 3 implementation of the batch GCD algorithms, however this can certainly be further optimized. First, by using a faster compiled language such as Rust, and second by parallelizing the algorithms. The remainder tree batch GCD algorithm has had parallelized implementations realized, and the batch GCD algorithm proposed in this study can also be parallelized very similarly. First, the product tree computations can be parallelized (with the added steps of GCD operations at each logical node in the binary tree for Step 1 of the algorithm). Second, the product tree computation can be parallelized when computing the aggregate shared factor integer $B$. Third, the GCD enumeration between each modulus and $B$ (Step 3 of the algorithm) can be embarrassingly parallelized.


\setlength\bibitemsep{0pt}
\printbibliography